\documentclass[showpacs,twocolumn,typeset,superscriptaddress]{revtex4}
\usepackage{amsfonts}
\usepackage{amsmath}
\usepackage{amssymb}
\usepackage{graphicx}
\usepackage{dcolumn}
\usepackage{bm}

\begin{document}
\preprint{APS}

\title{Entanglement thresholds for displaying the quantum nature of teleportation}

\author{Luis Roa}\email{lroa@udec.cl}
\affiliation{Departamento de F\'{\i}sica, Universidad de Concepci\'{o}n, Casilla 160-C, Concepci\'{o}n, Chile.}
\author{Robinson G\'omez}
\affiliation{Departamento de F\'{\i}sica, Universidad de Concepci\'{o}n, Casilla 160-C, Concepci\'{o}n, Chile.}
\author{Ariana Mu\~noz}
\affiliation{Departamento de F\'{\i}sica, Universidad de Concepci\'{o}n, Casilla 160-C, Concepci\'{o}n, Chile.}
\author{Gautam Rai}
\affiliation{Department of Physics \& Earth Sciences, Jacobs University Bremen, Germany.}

\date{\today}

\begin{abstract}
A protocol for transferring an unknown single qubit state has quantum features when the average fidelity of the outcomes is greater than $2/3$.
We use the probabilistic and unambiguous state extraction scheme as a mechanism to redistribute the fidelity in the outcome of the teleportation when the process is performed with an $X-$state as a noisy quantum channel.
We show that the entanglement of the channel is necessary but not sufficient in order for the total average fidelity $f_{\small X}$ to display quantum features, i.e., we find a threshold $C_{\small X}$ for the concurrence of the channel.
If the mechanism for redistributing fidelity is successfully applied then we find a filtrable outcome with normalized average fidelity $f_{\small X,USE,0}$ greater than $f_{\small X}$.
In addition, we find the threshold concurrence of the channel $C_{\small X,USE,0}$ in order for the normalized average fidelity to display quantum features.
Surprisingly, we find that the threshold concurrence $C_{\small X,USE,0}$ can be lesser than $C_{\small X}$.
Finally, we show that if the mechanism for redistributing fidelity fails then the respective filtrable outcome has average fidelity lesser than $2/3$.
\end{abstract}

 \maketitle

\section{Introduction}

Deterministic teleportation demands three important ingredients: a maximally entangled state as a quantum channel,
one-way transfer of classical information, and a pre-existing
agreement on the basis of measurement between the two spatially separated laboratories.
The quantum features of the channel are key resources, which can
be irreversibly affected by the interaction with the surrounding environment.
It is well known that different kinds of decoherence, dephasing, and dissipation mechanisms reduce purity and
entanglement of the channel.
In particular, it has been shown that in different physical systems, an initial Bell state evolves into a stationary $X$-state when it undergoes non-Markovian interaction mechanisms.
\cite{Rosario,Bellono}.
Therefore, an $X-$state is a good candidate to be considered as a real noisy quantum channel for implementing quantum protocols.
In this context much effort has been devoted to engineer those undesired effects in order to minimize the degrading effects \cite{Breuer}.
Similarly, a great interest has emerged in the scientific community to characterize the physical bounds and the kind of states, which, despite their mixture degree, are able to manifest quantum features.
In this context N. Gisin found the value $2/3$ as the upper bound for the fidelity when teleportation is performed with a particular separable state \cite{Gisin}.
S. Popescu shows that, in general, an average fidelity larger than $2/3$ appears when the channel has non-classical features \cite{Popescu}.
Additionally, a relation between the maximal fidelity of teleportation and the maximal
singlet fraction attainable by means of local quantum and classical communication action has been found \cite{Horodecki,BS}.
On the other hand, teleportation with a pure partially entangled state can be performed
probabilistically. In this case, the original teleporation scheme \cite{Bennett}
is complemented with an unambiguous state discrimination scheme \cite{Roa,Banaszek} or with an unambiguous state extraction
scheme \cite{Lis}.
Both schemes have the same probability of achieving successful teleportation, i.e., an outcome with fidelity $1$.
Recently, an interesting proposal shows a probabilistic scheme without losing the quantum information to be teleported \cite{RG}.

In this paper we show that the unambiguous state extraction (USE) scheme can be used to
redistribute the average fidelity in the outcome states when teleportation is performed with an $X-$state.
Additionally, we show that, in both processes, with and without USE, there is a competition between entanglement which introduces
quantum features and the dissipation mechanism which decreases the
population inside the principal subspace of the quantum channel.
Thus, we find threshold values of the channel concurrence in order for the quantum
features of the processes to be manifested in the average fidelity.
The paper is organized as follows:
In Sec. \ref{sec2} we describe briefly the object of analysis of this work, the average fidelity.
In Sec. \ref{sec3}, we review the
teleportation protocol with a pure quantum channel in order to characterize its main aspects.
In Sec. \ref{sec4}, we present a detailed analysis of teleportation carried out with an $X$-state as the quantum channel.
Finally, in the last section we summarize our principal results.

\section{Average Fidelity} \label{sec2}

Uhlmann's expression for transition probability $F=(Tr\sqrt{\sqrt{\rho}\varrho\sqrt{\rho}})^{2}$ was proposed as a well
defined fidelity of a quantum state $\rho $ with respect to another quantum state $\varrho $, or vice verse \cite{RJozsa}.
This fidelity accounts for how close or similar the states $\varrho $ and $\rho $ are; for instance, if $\rho =\varrho $ then $F=1$,
otherwise if $\rho \neq\varrho $ then $0\leq F<1$.
When the states belong to a two-dimensional Hilbert space, R.~Jozsa found that $F=Tr\left( \rho \varrho\right) +2\sqrt{\det \left( \rho \right) \det \left( \varrho \right) }$.
By noting that in this case $\det \left( \rho \right) =(1-Tr\rho ^{2})/2$ we obtain
\[
F=Tr\left( \rho \varrho \right) +\sqrt{(1-Tr\rho ^{2})(1-Tr\varrho ^{2})}.
\]
If at least one of the two states is pure, say $\rho =|\psi \rangle \langle \psi |$, then $F=\langle \psi |\varrho |\psi \rangle $.
In quantum information theory, the challenge is to transmit via a particular protocol an unknown pure state $|\psi \rangle $ with the highest fidelity possible.
However, due to some undesired process the received state can be different, say $\rho $ instead of $|\psi \rangle $.
Considering total ignorance of the emitted state $|\psi \rangle $, we assume that it could be any state of the
Hilbert space with the same probability density $\wp =1/2\pi ^{2}$.
Accordingly, the average fidelity $f$ is given by $F$ averaged onto all the possible states $\left\{ |\psi \rangle \right\} $ and it becomes
\begin{eqnarray*}
f &=&\int \wp Fd\psi , \\
&=&\frac{1}{2\pi ^{2}}\int_{0}^{2\pi }\int_{0}^{2\pi }\int_{0}^{1}\langle
\psi |\rho |\psi \rangle \left\vert \langle 0|\psi \rangle \right\vert
d\left\vert \langle 0|\psi \rangle \right\vert d\theta _{0}d\theta _{1},
\end{eqnarray*}%
where the $\theta_{i}$s are the two phases of the state $|\psi\rangle$, i.e. $\langle 0|\psi\rangle=\left\vert\langle 0|\psi\rangle\right\vert e^{i\theta_{0}}$ and
$\langle 1|\psi\rangle=\left\vert\langle 1|\psi\rangle\right\vert e^{i\theta_1}$, and $d\psi=\left\vert\langle 0|\psi\rangle\right\vert d\left\vert\langle 0|\psi\rangle\right\vert d\theta_{2}d\theta_{1}$
is an infinitesimal volume element in the space parameterized by the three independent variables $(\theta_{0},\theta_{1},\left\vert\langle 0|\psi\rangle\right\vert)$,
which define any state, and $\{|0\rangle,|1\rangle\}$ is any fixed orthogonal basis, henceforth the eigenstates of $\sigma_z$ Pauli operator.

Suppose that the emitter has a single system and wishes to transfer its unknown state $|\psi \rangle $ to a receiver.
The emitter has only a one-way classical channel for giving information to the receiver.
In this purely classical process the sender can measure an observable, whose eigenstates are $|0\rangle$ and $|1\rangle$, and send by means of a
classical bit a $0$ if the result is $|0\rangle$ or a $1$ if it is $|1\rangle$.
The observable has been previously agreed between them.
Thus, the receiver prepares the state $|0\rangle$ with probability $\left\vert\langle 0|\psi\rangle\right\vert^{2}$ and $|1\rangle$
with probability $\left\vert\langle 1|\psi\rangle\right\vert^{2}$, instead of $|\psi\rangle$.
In this case, the average fidelity of the state prepared by the receiver is given by
\begin{eqnarray}
f &=&\int \wp \langle \psi |\left( \left\vert \langle 0|\psi \rangle
\right\vert ^{2}|0\rangle \langle 0|+\left\vert \langle 1|\psi \rangle
\right\vert ^{2}|1\rangle \langle 1|\right) |\psi \rangle d\psi ,  \nonumber
\\
&=&\frac{2}{3},   \label{fmin}
\end{eqnarray}%
which means that the value $2/3$ is guaranteed by a purely classical process of transferring information.
In consequence, a process for transferring information has quantum features if the average fidelity of the outcome is
higher than $2/3$ \cite{Gisin,Popescu,Horodecki}.

\section{Redistributing fidelity in a process with a pure channel} \label{sec3}

Now we review succinctly the teleportation protocol with the pure quantum channel $|C_{AB}\rangle=\alpha|0\rangle|0\rangle+\beta|1\rangle
|1\rangle $, $|i\rangle |j\rangle $ denotes the tensorial product of the state $|i\rangle $ of system $A$ and state $|j\rangle$ of system $B$.
We assume that the systems $A$ and $B$ are spatially separated in such a way that joint operations between them can not be applied.
Since any phase in the channel can be removed by a local unitary operation, without loss of generality, we consider the
amplitudes $\alpha$ and $\beta$ to be non negative real numbers and $\alpha\leq\beta$.
The channel $|C_{AB}\rangle$ is normalized, so $\alpha^{2}+\beta^{2}=1$.
The entanglement of this channel can be characterized by its quantum concurrence, which is given by $C=\left\vert\langle\psi|\tilde{\psi}\rangle\right\vert=2\alpha\beta$ \cite{Wootters}.

The probabilistic teleportation \cite{Lis}, of an unknown state $|\psi \rangle $ of the qubit $a$, can be read from the following identity,
\begin{eqnarray}
|\psi \rangle |C_{AB}\rangle\!&=&\!\sqrt{\frac{\bar{p}}{2}}|\Phi
_{aA}^{+}\rangle |\bar{\psi}\rangle +\sqrt{\frac{\bar{p}}{2}}|\Phi
_{aA}^{-}\rangle \sigma _{z}|\bar{\psi}\rangle   \nonumber \\
&&+\sqrt{\frac{\ddot{p}}{2}}|\Psi _{aA}^{+}\rangle \sigma _{x}|\ddot{\psi}%
\rangle\!+\!\sqrt{\frac{\ddot{p}}{2}}|\Psi _{aA}^{-}\rangle \sigma _{x}\sigma
_{z}|\ddot{\psi}\rangle ,  \label{eq1}
\end{eqnarray}
where $|\Phi_{aA}^{\pm}\rangle=(|0\rangle|0\rangle\pm|1\rangle|1\rangle)/\sqrt{2}$ and $|\Psi_{aA}^{\pm}\rangle=(|0\rangle|1\rangle\pm|1\rangle|0\rangle)/\sqrt{2}$
are the Bell states of the bipartite system $aA$.
From Eq. (\ref{eq1}) one realizes that by performing a projective measurement onto the Bell states of the bipartite system $aA$,
the system $B$ is also projected onto the outcomes: $|\bar{\psi}\rangle $ with probability $\bar{p}/2$, $\sigma _{z}|\bar{\psi}\rangle $
with $\bar{p}/2$, $\sigma _{x}|\ddot{\psi}\rangle $ with $\ddot{p}/2$, and $\sigma _{x}\sigma _{z}|\ddot{\psi}\rangle $ with $\ddot{p}/2$, where
\begin{eqnarray}
|\bar{\psi}\rangle  &=&\frac{1}{\sqrt{\bar{p}}}\left( \alpha \langle 0|\psi
\rangle |0\rangle +\beta \langle 1|\psi \rangle |1\rangle \right) ,
 \label{sta1} \\
\bar{p} &=&\alpha ^{2}\left\vert \langle 0|\psi \rangle \right\vert
^{2}+\beta ^{2}\left\vert \langle 1|\psi \rangle \right\vert ^{2},  \nonumber
\\
|\ddot{\psi}\rangle  &=&\frac{1}{\sqrt{\ddot{p}}}\left( \beta \langle 0|\psi
\rangle |0\rangle +\alpha \langle 1|\psi \rangle |1\rangle \right) ,
  \label{est2} \\
\ddot{p} &=&1-\bar{p}.  \nonumber
\end{eqnarray}%
Once the receiver knows the measurement result, the unitary $\sigma _{z}$, $\sigma _{x}$ or $\sigma _{x}\sigma _{z}$ can be removed in order to retrieve
the states $|\bar{\psi}\rangle $ or $|\ddot{\psi}\rangle $ in system $B$.
It is worth mentioning that if $\beta =\left\vert \beta \right\vert e^{i\theta }$ is complex then we must introduce the unitary
$e^{-i\left(\theta +\pi \right) /2}e^{i\left( \theta +\pi \right) \sigma _{z}/2}$ instead of $\sigma _{z}$ in Eq. (\ref{eq1}).
Note that, when $\alpha =\beta$, the states $|\bar{\psi}\rangle$ and $|\ddot{\psi}\rangle$ become $|\psi\rangle$, which is the state to be teleported from qubit $a$ to qubit $B$.
Thus, after removing the unitaries, the receiver has only two outcomes, $|\bar{\psi}\rangle $ with probability $\bar{p}$ and $|\ddot{\psi}\rangle$ with probability $\ddot{p}$.
In this case, the average fidelity becomes,
\begin{eqnarray}
f_{p} &=&\int \wp \left( \bar{p}\left\vert \langle \psi |\bar{\psi}\rangle
\right\vert ^{2}+\ddot{p}\left\vert \langle \psi |\ddot{\psi}\rangle
\right\vert ^{2}\right) d\psi ,  \label{F11} \\
&=&\frac{2}{3}+\frac{C}{3}.
\end{eqnarray}%
From this expression we see that the entanglement correlation contributes to the upper third in the average fidelity since, as we saw,
the lower two thirds are guaranteed without any quantum correlation but with one-way classical communication.
In other words, in this case, only an entanglement different from zero gives the process a quantum nature.
Additionally, since each outcome $|\bar{\psi}\rangle $ and $|\ddot{\psi}\rangle$ can be filtered, we observe that both have the same normalized average fidelity,
$\bar{f}=\int\wp\bar{p}\left\vert \langle \psi |\bar{\psi}\rangle \right\vert ^{2}d\psi /\int \wp \bar{p}d\psi
=\ddot{f}=\int \wp \ddot{p}\left\vert \langle \psi |\ddot{\psi}\rangle \right\vert ^{2}d\psi /\int \wp \ddot{p}d\psi=f_{p}$, in consequence the filter is unnecessary.

Now, in order to redistribute the fidelity into each of the outcomes, $|\bar{\psi}\rangle $ and $|\ddot{\psi}\rangle $, we apply USE scheme \cite{Lis}.
For implementing USE we require an auxiliary qubit $b$ initially in $|0\rangle $.

If the outcome is $|\bar{\psi}\rangle $ then USE is carried out by applying $\bar{U}_{Bb}$ onto $|\bar{\psi}\rangle |0\rangle $, where
\begin{eqnarray*}
\bar{U}_{Bb} &=&|0\rangle \langle 0|\otimes I+|1\rangle \langle 1|\otimes U_{b}, \\
U_{b}|0\rangle  &=&\frac{\alpha }{\beta }|0\rangle -\sqrt{1-\frac{\alpha ^{2}%
}{\beta ^{2}}}|1\rangle , \\
U_{1}|1\rangle  &=&\sqrt{1-\frac{\alpha ^{2}}{\beta ^{2}}}|0\rangle +\frac{%
\alpha }{\beta }|1\rangle .
\end{eqnarray*}%
Thus, in this first stage of the USE process we have,
\begin{equation}
\bar{U}_{Bb}|\bar{\psi}\rangle |0\rangle =\frac{\alpha }{\sqrt{\bar{p}}}%
|\psi \rangle |0\rangle -\sqrt{\frac{\beta ^{2}-\alpha ^{2}}{\bar{p}}}%
\langle 1|\psi \rangle |1\rangle |1\rangle ,  \label{ESP1}
\end{equation}%
from where we realize that the unknown state $|\psi \rangle $ can be recovered by projecting the auxiliary qubit $b$ onto the state $|0\rangle $,
otherwise the system $B$ is projected onto the state $|1\rangle$.
The conditional probability of extracting $|\psi\rangle$, in this case, is $\alpha ^{2}/\bar{p}$.

If the outcome is $|\ddot{\psi}\rangle$ then we apply $\ddot{U}_{Bb}=\sigma_{x}^{(B)}\bar{U}_{Bb}\sigma_{x}^{(B)}$ onto $|\ddot{\psi}\rangle|0\rangle$.
In this case we have,
\begin{equation}
\ddot{U}_{Bb}|\ddot{\psi}\rangle |0\rangle =\frac{\alpha }{\sqrt{\ddot{p}}}%
|\psi \rangle |0\rangle -\sqrt{\frac{\beta ^{2}-\alpha ^{2}}{\ddot{p}}}%
\langle 0|\psi \rangle |0\rangle |1\rangle .   \label{EPS2}
\end{equation}%
Again the state $|\psi\rangle$ is recovered by projecting the auxiliary qubit $b$ onto the state $|0\rangle $,
otherwise the system $B$ is projected onto the state $|0\rangle$.
In this case the conditional probability of extracting $|\psi \rangle$ is $\alpha ^{2}/\ddot{p}$.
Accordingly, the total probability of extracting $|\psi \rangle $ from both outcomes becomes independent of $|\psi\rangle$ and it is given by
\begin{eqnarray*}
p_{ext} &=&\bar{p}\frac{\alpha ^{2}}{\bar{p}}+\ddot{p}\frac{\alpha ^{2}}{%
\ddot{p}} \\
&=&1-\sqrt{1-C^{2}}.
\end{eqnarray*}%
From this expression we see that the extraction of $|\psi\rangle $ is possible only if the concurrence $C$ of the channel is different from zero.
Both outcomes contribute to successful teleportation with the same probability $p_{ext}/2$.
However, when USE fails there are two possible outcomes, $|1\rangle$ or $|0\rangle$.
Thus, in this case, the average fidelity is given by
\begin{eqnarray*}
f_{p,{\small USE}}\!&\!=\!&\!\int \wp d\psi \left\{ \bar{p}\left[ \frac{\alpha ^{2}%
}{\bar{p}}\left\vert \langle \psi |\psi \rangle \right\vert ^{2}+\left( 1-%
\frac{\alpha ^{2}}{\bar{p}}\right) \left\vert \langle \psi |1\rangle
\right\vert ^{2}\right] \right.  \\
&&+\left. \ddot{p}\left[ \frac{\alpha ^{2}}{\ddot{p}}\left\vert \langle \psi
|\psi \rangle \right\vert ^{2}+\left( 1-\frac{\alpha ^{2}}{\ddot{p}}\right)
\left\vert \langle \psi |0\rangle \right\vert ^{2}\right] \right\} , \\
&=&1-\frac{\sqrt{1-C^{2}}}{3}.
\end{eqnarray*}%
We note that $f_{p,{\small USE}}$ reaches its minimum value $2/3$ when $C=0$ and its maximum $1$ for $C=1$.
Again, entanglement contributes to the upper third of the average fidelity.
Additionally, note that $f_{p,{\small USE}}$ is always lesser than $f_p$.
However, the process with USE allows us to obtain, with probability $p_{ext}$, the state $|\psi\rangle$, thus redistributing, in similar form, the fidelity into each of the two outcomes.

Here it is important to take into account that, by knowing the results of the measurements, the receiver can unambiguously filter the outcomes with the highest average fidelity, $1$ in this case.

These results confirm the claim at the end of Sec. \ref{sec2},
because both processes with and without USE have a quantum nature only if entanglement is different from zero, which is a purely quantum correlation \cite{Wootters,GOS,LAA}.
Finally, we note that in this case the entanglement of the channel is necessary and sufficient in order for the average fidelity to exhibit quantum features.

\section{Redistributing fidelity in a scheme with a noisy channel} \label{sec4}

Let us now study the same protocol with an $X$-state as the quantum channel instead of a pure one.
It is worth noting that an $X$-state populates two orthogonal subspaces, $\mathcal{H}_{00,11}$ spanned by the
basis $\left\{ |0\rangle |0\rangle ,|1\rangle |1\rangle \right\}$ and $\mathcal{H}_{01,10}$ spanned by $\left\{ |0\rangle |1\rangle ,|1\rangle|0\rangle \right\} $;
the $X$ form in itself arises because it has zero coherence between the elements of those two subspaces.
Thus, we consider systems $A$ and $B$ that share the following state,
\begin{eqnarray}
\hat{\rho}_{AB} &=&\rho _{11}|0\rangle |0\rangle \langle 0|\langle 0|+\rho_{14}|0\rangle |0\rangle \langle 1|\langle 1| \nonumber\\
&&+\rho_{22}|0\rangle |1\rangle \langle 0|\langle 1|+\rho_{23}|0\rangle|1\rangle\langle 1|\langle 0| \nonumber\\
&&+\rho_{32}|1\rangle |0\rangle \langle 0|\langle 1|+\rho_{33}|1\rangle|0\rangle\langle 1|\langle 0| \nonumber\\
&&+\rho_{41}|1\rangle |1\rangle \langle 0|\langle 0|+\rho_{44}|1\rangle|1\rangle\langle 1|\langle 1|. \label{x}
\end{eqnarray}
Without losing generality we consider the off-diagonal elements $\rho _{14}$ and $\rho _{23}$ to be real and non negative numbers \cite{RMG}.
Since the quantum channel considered in Sec. \ref{sec2} is in the subspace $\mathcal{H}_{00,11}$, here we consider it as the principal subspace, i.e.,
the entanglement of $\hat{\rho}_{AB}$ is given by the concurrence
\[
C=\max \left\{ 0,C_{14}\right\} ,
\]%
where%
\[
C_{14}=2\left( \rho _{14}-\sqrt{\rho _{22}\rho _{33}}\right) .
\]%
In other words, we assume that $C_{14}$ is higher than or equal to $C_{23}=2\left( \rho _{23}-\sqrt{\rho _{11}\rho _{44}}\right) \leq 0$; thus
when $C_{14}$ is smaller than zero then the channel lacks entanglement and it is separable.
If $C_{14}>0$ then $C=C_{14}$ and the channel is entangled.
Besides, this assumption implies that the populations of the two subspaces
satisfies the following inequality \cite{RMG}
\[
\rho _{11}\rho _{44}>\rho _{22}\rho _{33}.
\]%
It is worth realizing that the case with a pure channel is obtained for
$\rho _{11}=\alpha ^{2}$, $\rho _{44}=\beta ^{2}$, $\rho _{14}=\rho
_{41}=\alpha \beta $, and $\rho _{22}=\rho _{33}=0$.
Accordingly, we can think that the initial pure channel $|C_{AB}\rangle$ evolves to the state (\ref{x}) undergoing decoherence, which principally decreases $\rho_{14}$,
and dissipation, which principally populates the subspace $\mathcal{H}_{01,10}$.

With this channel we apply the teleportation process described in Sec. \ref{sec3}.
Firstly, the bipartite system $aA$ is measured in such a way that it can be projected onto one of the four Bell states $\left\{|\phi_{aA}^{\pm}\rangle,|\psi_{aA}^{\pm}\rangle\right\}$,
then the system $B$ is also projected onto one of the four states $\bar{\rho}_{B}^{\pm }$ with probability $\bar{p}/2$ or $\ddot{\rho}_{B}^{\pm }$ with probability $\ddot{p}/2$,
which are given by
\begin{eqnarray*}
\bar{\rho}_{B}^{\pm } &=&\frac{\langle \phi _{aA}^{\pm }|\left( |\psi
\rangle \langle \psi |\otimes \hat{\rho}_{AB}\right) |\phi _{aA}^{\pm
}\rangle }{Tr\langle \phi _{aA}^{\pm }|\left( |\psi _{a}\rangle \langle \psi
_{a}|\otimes \hat{\rho}_{AB}\right) |\phi _{aA}^{\pm }\rangle }, \\
&=&\sigma _{z}^{\left( 1\mp 1\right) /2}\bar{\varrho}_{B}\sigma _{z}^{\left(
1\mp 1\right) /2},
\end{eqnarray*}%
\begin{eqnarray*}
\ddot{\rho}_{B}^{\pm } &=&\frac{\langle \psi _{aA}^{\pm }|\left( |\psi
\rangle \langle \psi |\otimes \hat{\rho}_{AB}\right) |\psi _{aA}^{\pm
}\rangle }{Tr\langle \psi _{aA}^{\pm }|\left( |\psi _{a}\rangle \langle \psi
_{a}|\otimes \hat{\rho}_{AB}\right) |\psi _{aA}^{\pm }\rangle }, \\
&=&\sigma _{z}^{\left( 1\mp 1\right) /2}\sigma _{x}\ddot{\varrho}_{B}\sigma
_{x}\sigma _{z}^{\left( 1\mp 1\right) /2},
\end{eqnarray*}%
\begin{eqnarray*}
\bar{p} &=&Tr\langle \phi _{aA}^{\pm }|\left( |\psi _{a}\rangle \langle \psi
_{a}|\otimes \hat{\rho}_{AB}\right) |\phi _{aA}^{\pm }\rangle , \\
&=&\left( \rho _{11}+\rho _{22}\right) \left\vert \langle 0|\psi _{a}\rangle
\right\vert ^{2}+\left( \rho _{33}+\rho _{44}\right) \left\vert \langle
1|\psi _{a}\rangle \right\vert ^{2}, \\
\ddot{p} &=&Tr\langle \psi _{aA}^{\pm }|\left( |\psi _{a}\rangle \langle
\psi _{a}|\otimes \hat{\rho}_{AB}\right) |\psi _{aA}^{\pm }\rangle , \\
&=&1-\bar{p}.
\end{eqnarray*}%
In the second stage, once the measurement result is known the unitaries are removed.
So the receiver has, with probability $\bar{p}$, the state
\begin{eqnarray}
\bar{\varrho}_{B}\!&\!=\!&\!\frac{1}{\bar{p}}\left[ \left( \rho _{11}\left\vert
\langle 0|\psi \rangle \right\vert ^{2}+\rho _{44}\left\vert \langle 1|\psi
\rangle \right\vert ^{2}\right) |\bar{\psi}\rangle \langle \bar{\psi}%
| \right.  \nonumber \\
&&-\left( \sqrt{\rho _{11}\rho _{44}}-\rho _{14}\right)( \langle
0|\psi \rangle \langle \psi |1\rangle |0\rangle \langle 1| \nonumber \\
&&+\langle 1|\psi\rangle \langle \psi |0\rangle |1\rangle \langle 0|)   \nonumber \\
&&+\rho _{33}\left\vert \langle 1|\psi \rangle \right\vert ^{2}|0\rangle
\langle 0|+\rho _{22}\left\vert \langle 0|\psi \rangle \right\vert
^{2}|1\rangle \langle 1|  \nonumber \\
&& +\rho _{32}\langle 1|\psi \rangle \langle \psi |0\rangle |0\rangle\!
\langle 1|\!+\!\rho _{23}\langle 0|\psi \rangle \langle \psi |1\rangle |1\rangle\!
\langle 0|\Big]\!,  \label{rholine}
\end{eqnarray}%
and with probability $\ddot{p}$ the state
\begin{eqnarray}
\ddot{\varrho}_{B}\!&\!=\!&\!\frac{1}{\ddot{p}}\left[ \left( \rho _{44}\left\vert
\langle 0|\psi \rangle \right\vert ^{2}+\rho _{11}\left\vert \langle 1|\psi
\rangle \right\vert ^{2}\right) |\ddot{\psi}\rangle \langle \ddot{\psi}%
|\right.   \nonumber \\
&&-\left( \sqrt{\rho _{11}\rho _{44}}-\rho _{14}\right)( \langle
1|\psi \rangle \langle \psi |0\rangle |1\rangle \langle 0|\nonumber \\
&&+\langle 0|\psi\rangle \langle \psi |1\rangle |0\rangle \langle 1|)   \nonumber \\
&&+\rho _{22}\left\vert \langle 1|\psi \rangle \right\vert ^{2}|0\rangle
\langle 0|+\rho _{23}\langle 1|\psi \rangle \langle \psi |0\rangle |0\rangle
\langle 1|  \nonumber \\
&&\left. +\rho _{32}\langle 0|\psi \rangle \langle \psi |1\rangle |1\rangle
\langle 0|\!+\!\rho _{33}\left\vert \langle 0|\psi \rangle \right\vert
^{2}|1\rangle\!\langle 1|\right]\!,  \label{rhodotdot}
\end{eqnarray}%
where now we have defined the pure states%
\[
|\bar{\psi}\rangle =\frac{\sqrt{\rho _{11}}\langle 0|\psi \rangle |0\rangle +%
\sqrt{\rho _{44}}\langle 1|\psi \rangle |1\rangle }{\sqrt{\rho
_{11}\left\vert \langle 0|\psi \rangle \right\vert ^{2}+\rho _{44}\left\vert
\langle 1|\psi \rangle \right\vert ^{2}}},
\]%
\[
|\ddot{\psi}\rangle =\frac{\sqrt{\rho _{44}}\langle 0|\psi \rangle |0\rangle
+\sqrt{\rho _{11}}\langle 1|\psi \rangle |1\rangle }{\sqrt{\rho
_{44}\left\vert \langle 0|\psi \rangle \right\vert ^{2}+\rho _{11}\left\vert
\langle 1|\psi \rangle \right\vert ^{2}}}.
\]%
We note that the pure states $|\bar{\psi}\rangle \langle \bar{\psi}|$ in Eq.
(\ref{rholine}) and $|\ddot{\psi}\rangle \langle \ddot{\psi}|$ in (\ref%
{rhodotdot}) are affected by decoherence inside the subspace $\mathcal{H}_{00,11}$ and by the fact that the subspace $\mathcal{H}_{01,10}$ is populated.
In this case the average fidelity becomes
\begin{eqnarray}
f_{\small X} &=&\int \wp d\psi \left( \bar{p}\langle \psi |\bar{\varrho}%
_{B}|\psi \rangle +\ddot{p}\langle \psi |\ddot{\varrho}_{B}|\psi \rangle
\right) ,  \nonumber \\
&=&\frac{2}{3}+\frac{1}{3}\left[ 2\rho _{14}-\left( \rho _{22}+\rho
_{33}\right) \right] ,  \nonumber \\
&=&\frac{2}{3}+\frac{C_{14}-\left( \sqrt{\rho _{22}}-\sqrt{\rho _{33}}%
\right) ^{2}}{3}, \label{fx}
\end{eqnarray}%
From this expression we realize that entanglement is necessary but not sufficient for obtaining an average fidelity higher than $2/3$.
Specifically, the process displays quantum features only when $C_{14}$ is greater than the
threshold value $C_{\small X,th}$ given by
\begin{equation}
C_{\small X,th}=\left(\sqrt{\rho _{22}}-\sqrt{\rho _{33}}\right) ^{2}.  \label{CX}
\end{equation}%
This threshold value is zero when the channel is exclusively inside of $\mathcal{H}_{00,11}$ or when $\rho _{22}=\rho _{33}$.
More importantly, we realize from Eq. (\ref{fx}) that there is a competition between entanglement which favours the quantum features and increases the average
fidelity, and the dissipation mechanism which populates $\mathcal{H}_{01,10}$ and decreases the average fidelity.

Notice that $f_{\small X}$ and $C_{\small X,th}$ do not depend on the off-diagonal elements $\rho _{23}$, but do depend on how the subspace $\mathcal{H}_{01,10}$ is populated.
Here we also obtain that the normalized average fidelities of the two outcomes are equal, i.e.,
$\bar{f}=\int\wp\bar{p}\langle\psi|\bar{\varrho}_{B}|\psi\rangle d\psi/\int\wp\bar{p}d\psi=\ddot{f}=\int\wp\ddot{p}\langle\psi|\ddot{\varrho}_{B}|\psi\rangle d\psi/\int\wp\ddot{p}d\psi=f_{\small X}$.
Therefore, in this case, the discrimination or filtration between the two outcomes is unnecessary.

Now let us apply the USE scheme described in Sec. (\ref{sec3}).
In this case, we must replace $\alpha /\beta $ by $\sqrt{\rho _{11}/\rho _{44}}$ in the $\bar{U}_{Bb}$ and $\ddot{U}_{Bb}$,
where $\rho _{11}<\rho _{44}$ follows from our earlier assumption that $\alpha <\beta $.

If the outcome is $\bar{\varrho}_{B}$ then we transform $\bar{\varrho}_{B}\otimes |0\rangle \langle 0|$ by means of $\bar{U}_{Bb}$.
After applying $\bar{U}_{Bb}$, the $\sigma _{z}$ observable of the auxiliary system $b$ is measured.
Thus we calculate only the two possible results,
\begin{eqnarray}
\bar{\varrho}_{B,0_{b}} &=&\frac{\langle 0_{b}|\bar{U}_{Bb}\varrho
_{B}^{\phi }\otimes |0\rangle \langle 0|\bar{U}_{Bb}^{\dagger }|0_{b}\rangle
}{Tr\langle 0_{b}|\bar{U}_{Bb}\varrho _{B}^{\phi }\otimes |0\rangle \langle
0|\bar{U}_{Bb}^{\dagger }|0_{b}\rangle },  \nonumber \\
&=&\frac{1}{p\bar{p}}\bigg[ \rho _{11}|\psi \rangle \langle \psi | \nonumber \\
&&-\sqrt{\frac{\rho_{11}}{\rho_{44}}}\left(\sqrt{\rho_{11}\rho_{44}}-\rho_{14}\right)(\langle0|\psi\rangle\langle\psi|1\rangle|0\rangle\langle 1| \nonumber \\
&&+\langle 1|\psi \rangle \langle \psi |0\rangle |1\rangle\langle 0|)   \nonumber \\
&&+\rho _{33}\left\vert \langle 1|\psi \rangle \right\vert ^{2}|0\rangle\langle 0| \nonumber \\
&&+\!\sqrt{\frac{\rho _{11}}{\rho _{44}}}\rho _{23}(\langle 1|\psi\rangle \langle \psi |0\rangle |0\rangle \langle 1|\!+\!\langle 0|\psi\rangle \langle \psi |1\rangle |1\rangle \langle 0|)\nonumber \\
&& +\frac{\rho _{11}}{\rho_{44}}\rho _{22}\left\vert \langle 0|\psi \rangle \right\vert ^{2}|1\rangle
\langle 1|\bigg] ,  \label{rholine0}
\end{eqnarray}%
with conditional probability
\[
p=\frac{1}{\bar{p}}\left( \rho _{11}+\rho _{33}\left\vert \langle 1|\psi
\rangle \right\vert ^{2}+\frac{\rho _{11}}{\rho _{44}}\rho _{22}\left\vert
\langle 0|\psi \rangle \right\vert ^{2}\right) ,
\]%
and
\begin{eqnarray*}
\bar{\varrho}_{B,1_{b}} &=&\frac{\langle 1_{b}|\bar{U}_{Bb}\varrho
_{B}^{\phi }\otimes |0\rangle \langle 0|\bar{U}_{Bb}^{\dagger }|1_{b}\rangle
}{Tr\langle 1_{b}|\bar{U}_{Bb}\varrho _{B}^{\phi }\otimes |0\rangle \langle
0|\bar{U}_{Bb}^{\dagger }|1_{b}\rangle }, \\
&=&|1\rangle \langle 1|,
\end{eqnarray*}%
with conditional probability $1-p$.
Similarly, if the outcome is $\ddot{\varrho}_{B}$ then $\ddot{\varrho}_{B}\otimes |0\rangle \langle 0|$ is
transformed with $\ddot{U}_{Bb}$.
Here we calculate only the two possible results of system $B$ which arise due to the measurement of the system $b$; the state
\begin{eqnarray}
\ddot{\varrho}_{B,0_{b}} &=&\frac{\langle 1_{b}|\ddot{U}_{Bb}\ddot{\varrho}%
_{B}\otimes |0\rangle \langle 0|\ddot{U}_{Bb}^{\dagger }|1_{b}\rangle }{%
Tr\langle 1_{b}|\ddot{U}_{Bb}\varrho _{B}^{\psi }\otimes |0\rangle \langle 0|%
\ddot{U}_{Bb}^{\dagger }|1_{b}\rangle },  \nonumber \\
&=&\frac{1}{q\ddot{p}}\Big[ \rho _{11}|\psi \rangle \langle \psi |
\nonumber \\
&&-\sqrt{\frac{\rho_{11}}{\rho_{44}}}\left(\sqrt{\rho_{11}\rho_{44}}-\rho_{14}\right)(\langle 0|\psi\rangle\langle\psi|1\rangle|0\rangle\langle 1| \nonumber \\
&&+\langle 1|\psi\rangle\langle\psi|0\rangle|1\rangle\langle 0|)   \nonumber \\
&&+\frac{\rho_{11}}{\rho_{44}}\rho_{22}\left\vert\langle 1|\psi\rangle\right\vert^{2}|0\rangle\langle 0| \nonumber \\
&&+\sqrt{\frac{\rho_{11}}{\rho_{44}}}\rho_{23}(\langle 1|\psi\rangle\langle\psi|0\rangle|0\rangle\langle 1|\!+\!\langle 0|\psi\rangle\langle\psi|1\rangle|1\rangle\langle 0|) \nonumber \\
&&\left.+\rho_{33}\left\vert\langle 0|\psi\rangle\right\vert^{2}|1\rangle\langle 1|\right],
\label{rhodotdot0}
\end{eqnarray}%
appears with conditional probability%
\[
q=\frac{1}{\ddot{p}}\left( \rho _{11}+\rho _{33}\left\vert \langle 0|\psi
\rangle \right\vert ^{2}+\frac{\rho _{11}}{\rho _{44}}\rho _{22}\left\vert
\langle 1|\psi \rangle \right\vert ^{2}\right) ,
\]%
and
\begin{eqnarray*}
\ddot{\varrho}_{B,1_{b}} &=&\frac{\langle 1_{b}|\ddot{U}_{Bb}\ddot{\varrho}%
_{B}\otimes |0\rangle \langle 0|\ddot{U}_{Bb}^{\dagger }|1_{b}\rangle }{%
Tr\langle 1_{b}|\ddot{U}_{Bb}\varrho _{B}^{\psi }\otimes |0\rangle \langle 0|%
\ddot{U}_{Bb}^{\dagger }|1_{b}\rangle } \\
&=&|0\rangle \langle 0|.
\end{eqnarray*}%
with conditional probability $1-q$.
Notice that, both outcome states (\ref{rholine0}) and (\ref{rhodotdot0}) are mixed states composed of the desired state $|\psi\rangle\langle\psi|$,
non-diagonal terms which appear because of the decoherence inside $\mathcal{H}_{00,11}$, and other terms coming from the population of subspace $\mathcal{H}_{01,10}$.
Thus, the extraction becomes approximate due to those undesired terms.
Therefore, the total probability of quasiextracting $|\psi \rangle \langle \psi |$ becomes
\begin{eqnarray*}
p_{\sim ext} &=&\bar{p}p+\ddot{p}q, \\
&=&2\rho _{11}+\rho _{33}+\frac{\rho _{11}\rho _{22}}{\rho _{44}},
\end{eqnarray*}%
and the total average fidelity of this process is given by
\begin{eqnarray*}
f_{\small X,USE} &=&\int \wp \left[ \bar{p}\left( p\langle \psi |\bar{%
\varrho}_{B,0_{b}}|\psi \rangle +\left( 1-p\right) \langle \psi |1\rangle
\langle 1|\psi \rangle \right) \right.  \\
&&+\left. \ddot{p}\left( q\langle \psi |\ddot{\varrho}_{B,0_{b}}|\psi
\rangle +\left( 1-q\right) \langle \psi |0\rangle \langle 0|\psi \rangle
\right) \right] d\psi , \\
&=&\frac{2}{3}\left[ 1+\rho _{14}\sqrt{\frac{\rho _{11}}{\rho _{44}}}-\frac{1%
}{2}\left( \rho _{22}+\rho _{33}\right) \right] , \\
&=&\frac{2}{3}+\frac{1}{3}\sqrt{\frac{\rho _{11}}{\rho _{44}}}\left[C_{14}-\left( \sqrt{\rho _{22}}-\sqrt{\rho _{33}}\right) ^{2} \right.\\
&&\left.-\left( \rho_{22}+\rho _{33}\right) \left( \sqrt{\frac{\rho _{44}}{\rho _{11}}}-1\right)\right].
\end{eqnarray*}%
Here we realize that there is a threshold value $C_{\small X,USE,th}$ of the concurrence in order to ensure that $f_{\small X,USE}$ is higher than $2/3$, where
\begin{equation*}
C_{\small X,USE,th}\!=\!\left( \sqrt{\rho _{22}}-\sqrt{\rho _{33}}\right) ^{2}+\left(\sqrt{\frac{\rho _{44}}{\rho _{11}}}-1\right) \left( \rho _{22}+\rho_{33}\right).
\end{equation*}
This shows that entanglement is necessary but not sufficient, i.e., the process displays quantum features only when
\[
C_{14}>C_{\small X,USE,th}.
\]
Note that $C_{\small X,USE,th}$ becomes zero when the populations inside both subspaces are uniform.
Clearly the threshold $C_{{\small X-USE},th}$ is greater than $C_{X,th}$, i.e., this procedure with USE demands more
entanglement in order to have total average fidelity higher than $2/3$.
Besides, we can easily note that $f_{\small X,USE}\leq f_{\small X}$.
Therefore we ask what do we gain by implementing the USE process?
For answering this question we have to take into account that the receiver can filter the outcomes of USE process.
In this form, the receiver can unambiguously choose the outcomes with the greatest normalized average fidelity.
The normalized average fidelity of the outcomes associated with the quasiextraction is
\begin{eqnarray*}
f_{\small X,USE,0}\!&\!=\!&\!\frac{1}{p_{\sim ext}}\!\int\!\!\wp(\bar{p}p\langle \psi |\bar{\varrho}_{B,0_{b}}|\psi \rangle\!+\!\ddot{p}q\langle \psi|\ddot{\varrho}_{B,0_{b}}|\psi \rangle\!) d\psi , \\
&=&\frac{2}{3}+\frac{2\sqrt{\frac{\rho _{11}}{\rho _{44}}}\rho _{14}-\left(
\rho _{33}+\frac{\rho _{11}}{\rho _{44}}\rho _{22}\right) }{3\left( 2\rho
_{11}+\rho _{33}+\frac{\rho _{11}\rho _{22}}{\rho _{44}}\right) }, \\
&=&\frac{2}{3}+\sqrt{\frac{\rho _{11}}{\rho _{44}}}\frac{%
C_{14}-C_{\small X,USE,0,th}}{3\left( 2\rho _{11}+\rho _{33}+\frac{\rho _{11}\rho _{22}}{\rho _{44}}\right) },
\end{eqnarray*}
which is higher than $2/3$ when the channel concurrence is higher than the threshold value $C_{\small X,USE,0,th}$ given by
\begin{eqnarray}
C_{\small X,USE,0,th} &=&\frac{\left( \sqrt{\rho _{11}\rho _{22}}-\sqrt{\rho
_{33}\rho _{44}}\right) ^{2}}{\sqrt{\rho _{11}\rho _{44}}},  \nonumber \\
\!&\!=\!&\!C_{\small X,USE,th}\!-\!\rho _{22}\!\left(\!\sqrt{\frac{\rho _{44}}{\rho _{11}}}\!-\!\sqrt{\frac{\rho _{11}}{\rho _{44}}}\!\right)\!.  \label{CX0}
\end{eqnarray}%
We note that in this case $f_{\small X,USE,0}>f_{\small X}$, which is what we gain.
Besides, we can see that the threshold value $C_{\small X,USE,0,th}$ is smaller than $C_{\small X,th}$ when
\[
\rho _{33}<\rho _{22}\sqrt{\frac{\rho _{11}}{\rho _{44}}},
\]%
which is another effect that we gain.
Note also that $C_{\small X,USE,0,th}$ becomes zero for $\rho _{11}\rho _{22}=\rho _{33}\rho _{44}$.
On the other hand, if the quasiextraction fails then the normalized average fidelity of the other outcomes becomes
\begin{eqnarray*}
f_{\small X,USE,1}&=&\frac{1}{1-p_{\sim ext}}\int\wp\left[\bar{p}(1-p)|\langle 1|\psi\rangle|^{2}\right.\\
&&\left.+\ddot{p}(1-q)|\langle0|\psi\rangle|^{2}\right]d\psi, \\
&=&\frac{2}{3}-\frac{\rho _{22}\left( 1-\frac{\rho _{11}}{\rho _{44}}\right)
}{3\left( \rho _{22}+\rho _{44}-\rho _{11}-\frac{\rho _{11}\rho _{22}}{\rho
_{44}}\right) }.
\end{eqnarray*}%
Clearly $f_{\small X,USE,1}$ is smaller than $2/3$, which is the price
that we pay.

Therefore, here the USE procedure allows us to redistribute the fidelity between two outcomes; one of them with average fidelity grater than $2/3$
if the concurrence of the channel is higher that the threshold (\ref{CX0}), and the other one always lesser that $2/3$.
Again we note from the expression of the threshold $C_{\small X,USE,th}$ and $C_{\small X,USE,0,th}$ that there is a competition between entanglement, which introduces the quantum features, and
the decoherence and dissipation mechanisms, which introduce classicality to the quantum channel.

\section{Conclusion}  \label{sec5}

In summary we have proposed a scheme that redistributes the average fidelity in a teleportation procedure, with a noisy channel modelled by an $X-$state.
We make use of the unambiguous state extraction protocol to perform the redistribution of the fidelity inside each outcome state of the teleportation process.
The fidelity redistribution allows us to enhance significantly the manifestation of the quantum nature in a distinguishable outcome with probability different from zero.

Specifically, we find that when the quantum channel is pure, the entanglement of the channel is necessary and sufficient in order for the quantum nature to emerge in the total average fidelity of
the outcome states. In this case, the USE redistributes the average fidelity in such a way that there is a possible outcome with fidelity $1$, which means that the teleportation is successful.

When the quantum channel is an $X$-state, we find that the entanglement of the channel is necessary but not sufficient for having a total average fidelity greater than $2/3$.
In this case we have found a threshold value of the concurrence, beyond which the quantum nature of teleportation emerges in the total average fidelity.
By implementing the USE scheme the fidelity is redistributed in a way such that there is a filtrable outcome, with probability different from zero, with average fidelity significantly higher than that without the USE.
Furthermore, the threshold concurrence for getting this effect can be smaller than that for the process without the USE.

Finally we want to emphasize that when the quantum channel is a mixed $X$-state, we easily see that there appears a competition between entanglement
which introduces a quantum nature and the decoherence and dissipation mechanisms which introduce classicality.

Authors thank grants FONDECyT 1120695, Conicyt-PCHA/DocNac/2014-21140554, and DAAD RISE Worldwide CL-PH-1170.

%\bibliography{references}

\begin{thebibliography}{99}

\bibitem{Bennett} C. Bennett, G. Brassard, C. Crepeau, R. Jozsa, A. Peres, and W.
K. Wootters, Teleporting an unknown
quantum state via dual classical and Einstein-Podolsky-Rosen channels, Phys. Rev. Lett. \textbf{70}, 1895 (1993);
D. BouwmeesterJ.-W. Pan, K. Mattle, M. Eibl, H. Weinfurter, and A.
Zeilinger, Experimental quantum teleportation, Nature (London) \textbf{390}, 575 (1997);
D. Boschi, S. Brance, F. De Martini, L. Hardy, and S. Popescu, Experimental Realization of Teleporting an Unknown
Pure Quantum State via Dual Classical and Einstein-Podolsky-Rosen Channels, Phys. Rev.
Lett. \textbf{80}, 1121 (1998).
\bibitem{Rosario} T. Yu and J. H. Eberly, Quantum Open System Theory: Bipartite Aspects, Phys. Rev. Lett. \textbf{97}, 140403 (2006);
M. F. Santos, P. Milman, L. Davidovich, and N. Zagury, Direct measurement of finite-time disentanglement induced
by a reservoir, Phys. Rev. A. \textbf{73}, 040305(R) (2006);
Qi-liang He and Jing-bo Xu, Sudden
transition and sudden change of quantum discord in dissipative cavity
quantum electrodynamics system, J. Opt. Soc. Am. B \textbf{30}, 251 (2013);
F. Lastra, C. E. Lopez, L. Roa, and J. C. Retamal, Entanglement of formation for a family of (2$\otimes$d)-dimensional
systems, Phys. Rev. A \textbf{85}, 022320 (2012);
L. Roa, A. Krugel, and C. Saavedra, Quantum
state stability against decoherence, Phys. Lett. A \textbf{366}, 563 (2007);
R. Lo Franco, B. Bellomo, S. Maniscalco, and G. Compagno, Dynamics of quantum correlations in two-qubit systems
within non-markovian environments, Int. J. Mod. Phys. B \textbf{27}, 1345053 (2013);
B. Bellomo, G. Compagno, A. D'Arrigo, G. Falci, R. Lo Franco, and E.
Paladino, Entanglement degradation in the
solid state: Interplay of adiabatic and quantum noise, Phys. Rev. A \textbf{81}, 062309 (2010);
R. Lo Franco, A. D'Arrigo, G. Falci, G. Compagno, and E. Paladino, Entanglement dynamics in superconducting qubits
affected by local bistable impurities, Phys. Scr. \textbf{T147}, 014019 (2012).
\bibitem{Bellono} B. Bellomo, G. Compagno, R. Lo Franco, A. Ridolfo, and S. Savasta, Dynamics and extraction of quantum discord in a
multipartite open system, Int. J.
Quantum Inf. \textbf{9}, 1665 (2011);
B. Aaronson, R. Lo Franco, and G. Adesso, Comparative investigation of the freezing phenomena for quantum
correlations under nondissipative decoherence, Phys. Rev. A \textbf{88}, 012120 (2013);
B. Aaronson, R. Lo Franco, G. Compagno, and G. Adesso, Hierarchy and dynamics of trace distance correlations, New J. Phys. \textbf{15},
093022 (2013);
B. Bellomo, R. Lo Franco, S. Maniscalco, and G. Compagno, Two-qubit entanglement dynamics for two different
non-Markovian environments, Phys. Scr. \textbf{T140},
014014 (2010).
\bibitem{Breuer} H.-P. Breuer and F. Petruccione, The Theory of Open Quantum Systems (Oxford University Press, Oxford, UK, 2007).
\bibitem{Gisin} N. Gisin, Nonlocality criteria
for quantum teleportation, Phys, Lett. A \textbf{210}, 157 (1996).
\bibitem{Popescu} S. Popescu, Bell's inequalities
versus teleportation: What is nonlocality?, Phys. Rev. Lett. \textbf{72}, 797 (1994).
\bibitem{Horodecki} M. Horodecki, P. Horodecki, and R. Horodecki, General teleportation channel, singlet fraction, and
quasidistillation, Phys. Rev. A \textbf{60}, 1888 (2013);
F. Verstraete and H. Verschelde, Optimal
Teleportation with a Mixed State of Two Qubits, Phys. Rev. Lett. \textbf{90}, 097901 (2003).
\bibitem{BS} S. Bandyopadhyay and B.C. Sanders, Quantum teleportation of composite systems via mixed entangled states, Phys. Rev. A \textbf{74}, 032310 (2006).
\bibitem{Roa} L. Roa, A. Delgado, and I. Fuentes-Guridi, Optimal conclusive teleportation of quantum states, Phys. Rev. A \textbf{68}, 022310
(2003).
\bibitem{Banaszek} K. Banaszek, Optimal quantum
teleportation with an arbitrary pure state, Phys. Rev. A \textbf{62}, 024301 (2000);
\bibitem{Lis} P. Agrawal, A. K. Pati, Probabilistic quantum teleportation, Phys. Letts. A, \textbf{305}, 12 (2002);
Li-Yi Hsu, Optimal information extraction in
probabilistic teleportation, Phys. Rev. A \textbf{66}, 012308 (2002);
Wan-Li Li, Chuan-Feng Li, and Guang-Can Guo, Probabilistic teleportation and entanglement matching, Phys. Rev. A \textbf{61}, 034301 (2000).
\bibitem{RG} L. Roa and C. Groiseau, Probabilistic
teleportation without loss of information, Phys. Rev. A \textbf{91}, 012344 (2015).
\bibitem{RJozsa} R. Jozsa, Fidelity for
Mixed Quantum States, J. of Mod. Optics \textbf{41}, 2315 (1994);
R. Jozsa and B. Schumacher, A new proof of
the quantum noiseless coding theorem, J. mod. Optics, \textbf{41}, 2343 (1994);
A. Uhlmann, The transition probability in the state space of a *-algebra, Rep. Math. Phys. \textbf{9}, 273 (1976).
\bibitem{Wootters} W. K. Wootters, Entanglement
of Formation of an Arbitrary State of Two Qubits, Phys. Rev. Lett. \textbf{80}, 2245 (1998);
S. Hill and W. K. Wootters, Entanglement of a Pair of Quantum Bits, ibid. \textbf{78}, 5022 (1997).
\bibitem{GOS} V. Gheorghiu, M.C. de Oliveira, and B.C. Sanders, Nonzero Classical Discord, Phys. Rev. Lett. \textbf{115}, 030403 (2015).
\bibitem{LAA} K. Modi, T. Paterek, W. Son, V. Vedral, and M. Williamson, Unified View of Quantum and Classical
Correlations, Phys. Rev. Lett. \textbf{104}, 080501 (2010);
L. Roa, A. Maldonado-Trapp, and A. B. Klimov, A
measure for maximum similarity between outcome states, EPL \textbf{109}, 400001 (2015);
L. Roa, J. C. Retamal, and M. Alid-Vaccarezza, Dissonance is Required for Assisted Optimal State Discrimination, Phys. Rev. Lett. \textbf{107}, 080401
(2011);
L. Henderson and V. Vedral, Classical, quantum and total correlations, J. Phys. A: Math. Gen. \textbf{34}, 6899 (2001);
H. Ollivier and W. H. Zurek, Quantum
Discord: A Measure of the Quantumness of Correlations, Phys. Rev. Lett. \textbf{88}, 017901 (2001).
\bibitem{RMG} L. Roa, A. Mu\~noz, and G. Gr\"uning, Entanglement swapping for $X$ states demands threshold values, Phys. Rev. A \textbf{89}, 064301 (2014).
\end{thebibliography}
%\bibliographystyle{unsrt}
%\addcontentsline{toc}{Section}{References}

\end{document}